\documentclass[final]{iopart}
\usepackage[final]{graphicx}
\usepackage{epsfig}
\usepackage{color}
\usepackage{ulem}

\begin{document}

\title[Uncovering selective excitations using RIXS in correlated materials]{Uncovering selective excitations using the resonant profile of indirect inelastic x-ray scattering in correlated materials: Observing two-magnon scattering and relation to the dynamical structure factor}

\author{C.~J.~Jia$^{1,2}$, C.-C.~Chen$^{1}$, A.~P.~Sorini$^{3}$, B. Moritz$^{1,4}$, T. P. Devereaux$^{1,5}$}
\address{$^1$SIMES, SLAC National Accelerator Laboratory, Menlo Park, California 94025, USA}
\address{$^2$Department of Applied Physics, Stanford University, Stanford, California 94305, USA}
\address{$^3$Lawrence Livermore National Laboratory, Livermore, CA 94550, USA}
\address{$^4$Department of Physics and Astrophysics, University of North Dakota, Grand Forks, ND 58202, USA}
\address{$^5$Geballe Laboratory for Advanced Materials, Stanford University, Stanford, California 94305, USA}

\begin{abstract}
Resonant inelastic x-ray scattering (RIXS) is a spectroscopic technique which has been widely used to study various elementary excitations in correlated and other condensed matter systems. For strongly correlated materials, besides boosting the overall signal the dependence of the resonant profile on incident photon energy is still not fully understood. Previous endeavors in connecting indirect RIXS, such as Cu \textit{K}-edge for example where scattering takes place only via the core-hole created as an intermediate state, with the charge dynamical structure factor $S(q,\omega)$ neglected complicated dependence on the intermediate state configuration. To resolve this issue, we performed an exact diagonalization study of the RIXS cross-section using the single-band Hubbard model by fully addressing the intermediate state contribution. Our results are relevant to indirect RIXS in correlated materials, such as high Tc cuprates. We demonstrate that RIXS spectra can be reduced to  $S(q,\omega)$ when there is no screening channel for the core-hole potential in the intermediate state. We also show that two-magnon excitations are highlighted at the resonant photon energy when the core-hole potential in the corresponding intermediate state is poorly screened. Our results demonstrate that different elementary excitations can be emphasized at different intermediate states, such that selecting the exact incident energy is critical when trying to capture a particular elementary excitation.
\end{abstract}

\pacs{78.70.Ck, 74.72 -h, 78.20 Bh}

\date{\today}
\maketitle

\section{Introduction}

Resonant inelastic x-ray scattering (RIXS) is a widely used technique for studying various condensed matter materials, including semiconductors, metals, oxides, and transition-metal compounds\cite{KotaniReview, TomRIXSReview}. Because of its many advantages, including bulk sensitivity, orbital and elemental specificity and polarization control, RIXS has moved to the forefront of important spectroscopic techniques in recent years\cite{NatPhyRIXS}. While in general, the probability of x-rays to be scattered from a solid is small, the cross section can be enhanced by orders of magnitude when tuning the incoming x-ray energy to a resonance of the system where an electron from a deep lying core state is excited into the
valence shell. The resulting intermediate core-hole state can connect the many-body ground and final excited states through pathways that may be inaccessible using other non-resonant x-ray probes, thus providing a wealth of new information about charge, spin, lattice and orbital degrees of freedom. This is particularly important in strongly correlated transition-metal oxides where these degrees of freedom are heavily intertwined. In this way, RIXS can offer complementary investigation of elementary excitations along with resonant diffraction, neutron scattering, angle resolved photoemission spectroscopy(ARPES) and scanning tunneling microscopy. In particular, as resonant diffraction yields understanding of the ground state of correlated materials, we show that RIXS can provide the same sort of understanding of excited state properties. By making explicit use of the resonance profile and associated pathways, one can understand simpler density response in even the most strongly correlated systems and the more complicated shake-up processes involved in RIXS.

RIXS spectra display both incident photon energy and energy loss dependence, where the underlying multi-particle process makes it hard to interpret the experimental cross-section\cite{Tom2007Review}. One of the main efforts to better understand this cross-section for correlated materials has been devoted to disentangling the resonant profile and study its response function, which only depends on the energy loss. As proposed under certain approximations involving intermediate states for indirect RIXS, (the case as the Cu \textit{K}-edge for example where scattering takes place only via the core-hole created as an intermediate state)\cite{AbbamonteRIXS,VDBRIXS,LuukRIXS}, the cross-section could be decomposed into a Lorentzian resonant profile and a response function, which is directly connected to the charge dynamical structure factor $S(\mathbf{q},\omega)$. Experimentally, in most of the endeavors to interpret the response function of RIXS spectra for correlated materials, either a particular incident photon energy has been considered\cite{YJKim2004,YJKim2007,Grenier} or the incident energy dependence was treated using this Lorentzian profile derived from theoretical approximations\cite{AbbamonteRIXS}. However, these experiments do not provide definitive proof that the response function can be directly related to $S(\mathbf{q},\omega)$, and the validity of the above approximations still need to be discussed for particular cases. In particular, for the Cu $K$-edge measurements on various cooper oxides, RIXS cast in terms of the non-resonant response applies for some materials, but strong deviations were found for other materials\cite{YJKimRIXS}. This suggests, at least in correlated materials, that the validity of treating the RIXS response function as $S(\mathbf{q},\omega)$ needs to be carefully investigated. Furthermore, RIXS is known to display other elementary excitations for correlated materials, such as magnon excitations\cite{TwoMagnonEllis,TwoMagnonForte,TwoMagnonJohnHill}, which are beyond the simple charge response provided by $S(\mathbf{q},\omega)$. Thus, in concert with boosting the overall signal, the dependence of the resonant profile on incident photon energy is an important consideration in strongly correlated materials.

In this work, we performed an exact diagonalization study of the single-band Hubbard model to understand the indirect RIXS cross-section in strongly correlated materials. The intermediate state contribution has been fully considered, so that the incident energy dependent resonant profile could be investigated. This goes beyond previous theoretical treatments using approximations in which the complicated intermediate state dependence has been ignored\cite{AbbamonteRIXS,VDBRIXS,LuukRIXS}. The result can be connected to the Cu \textit{K}-edge RIXS for high Tc cuprates. In two examples, we show how the incident energy dependent RIXS response can be related to the charge response $S(\mathbf{q},\omega)$ and how the incident photon energy may be used to highlight two-magnon scattering. 

Our results show that the RIXS spectra display strong resonances at multiple incident photon energies, each of which corresponds to a different intermediate state electronic configuration. We find that only at particular intermediate states where the screening of the core-hole potential and correlation effect are not important, one observes a RIXS spectrum similar to $S(\mathbf{q},\omega)$. However, RIXS behaves very differently when compared to $S(\mathbf{q},\omega)$ at other incident energies where the screening of the core-hole potential is important in the corresponding intermediate state. Moreover, different elementary excitations are highlighted at different incident photon energies, in which, strong intensities for two-magnon excitations are found at the so called poorly-screened intermediate states. The two-magnon excitations are not shown at the intermediate state where the core-hole potential screening is absent. The significance of emphasized two-magnon excitation at one intermediate state compared to the others provides a focus to the fact that tuning to a proper incident photon energy dictated by the character of the intermediate state is critical to study the elementary excitations in RIXS.

\section{Methods}

We employ the single-band Hubbard Hamiltonian that serves as an effective low energy model for studying correlated materials, written as
\begin{equation}\label{Hubbard}
\mathit{H}= -\sum_{\mathbf{i},\mathbf{j},\sigma} t_{\mathbf{i},\mathbf{j}} d_{\mathbf{i},\sigma}^{\dagger}d_{\mathbf{j},\sigma}
+\sum_{\mathbf{i}} U n_{\mathbf{i},\uparrow}^d n_{\mathbf{i},\downarrow}^d,
\end{equation}
where $d_{\mathbf{i},\sigma}^{\dagger}$ ($d_{\mathbf{i},\sigma}$) creates (annihilates) a fermion with spin $\sigma$ on lattice site $\mathbf{i}$, $t_{\mathbf{i},\mathbf{j}}$ is the hopping integral restricted to the nearest- and next-nearest-neighbor, denoted as $t$ and $t^{\prime}$. $U$ is the on-site Coulomb repulsion, and $n_{\mathbf{i},\sigma}^d$ is the fermion number operator. 

The momentum-dependent indirect RIXS cross section can be expressed using the Kramers-Heisenberg formula\cite{KramersHeisenberg}
\begin{equation}\label{RIXS}
I(\mathbf{q}, \Omega, \omega_i)
= \frac{1}{\pi} \mathrm{Im} \langle \Psi | \frac{1}{\mathit{H} - \mathit{E}_0 -\Omega  - \mathit{i} 0^+} | \Psi \rangle
\end{equation}
and 
\begin{equation}\label{psi}
| \Psi \rangle = \sum_{\mathbf{i},\sigma} e^{i\mathbf{q}\cdot\mathbf{r}_{\mathbf{i}}} \mathit{D}^{\dagger} \frac{1}{\mathit{H}^{\prime}-\mathit{E}_0-\omega_i-\mathit{i}\Gamma} \mathit{D} | 0 \rangle,
\end{equation}
where $\mathbf{q}$ is the momentum transfer; $\omega_i$ and $\Omega = \omega_i - \omega_f$ are the incident energy and energy transfer, respectively; $\Gamma$ is the inverse core-hole lifetime; $\mathit{E}_0$ is the ground state energy of the system absent the core-hole; $| 0 \rangle$ is the ground state wave function; $\mathit{H}^{\prime}=\mathit{H}-\mathit{U}_c
\sum_{\mathbf{i}, \sigma, \sigma^{\prime}} n_{\mathbf{i}\sigma}^d n_{\mathbf{i}, \sigma^{\prime}}^s$ represents the intermediate state Hamiltonian, which includes the core-hole potential with strength $U_c$, with $n_{\mathbf{i},\sigma}^s$ denoting the core-hole number operator; $\mathit{D}$ is the dipole transition operator, which for example excites a $1s$ core-electron up to $4p$ level and is formulated as $\mathit{D} = p_{\mathbf{i},\sigma}^{\dagger}s_{\mathbf{i},\sigma}$ in transition-metal $K$-edge indirect RIXS. The wave function overlap of the intermediate state with the final/ground state is fully considered, and thus the whole resonant profile can be investigated using this method without further approximation. Elastic lines are trivial and have been removed in the results in the following sections.

We also investigate the x-ray absorption spectroscopy (XAS) given by
\begin{equation}\label{XAS}
B(\omega_i)= \frac{1}{\pi} \mathrm{Im} \langle 0 | \mathit{D}^{\dagger} \frac{1}{\mathit{H}^{\prime}-\mathit{E}_0-\omega_i-\mathit{i}\Gamma} \mathit{D} | 0 \rangle
\end{equation}
in which $\mathit{D}$ and $\mathit{H}^{\prime}$ are the same as in the RIXS's definition. The XAS peaks can be connected with the resonant energies in the RIXS spectra.

The charge dynamical structure factor to be compared to RIXS is
\begin{equation}\label{SQW}
S(\mathbf{q},\omega)= 
\frac{1}{\pi} \mathrm{Im} \langle 0 | \rho_{-\mathbf{q}}^d \frac{1}{H - \mathit{E}_0 - \omega - \mathit{i}0^+} \rho_{\mathbf{q}}^d | 0 \rangle,
\end{equation}
where $\rho_{\mathbf{q}}^d=\sum_{\mathbf{k},\sigma}d_{\mathbf{k+q},\sigma}^{\dagger} d_{\mathbf{k},\sigma}=\sum_{\mathbf{i},\sigma}e^{i\mathbf{q}\cdot \mathbf{R}_{\mathbf{i}}}n_{\mathbf{i},\sigma}^d$ is the density operator. We note that the charge dynamical structure factor $S(\mathbf{q},\omega)$, which represents the non-resonant response, is an effective two-particle process; however for resonant response the intermediate core-hole state increases the complexity of the problem to an effective four-particle process making its interpretation more difficult\cite{Tom2007Review}.

\begin{figure}[t]
\begin{center}
\includegraphics[width=0.8\columnwidth]{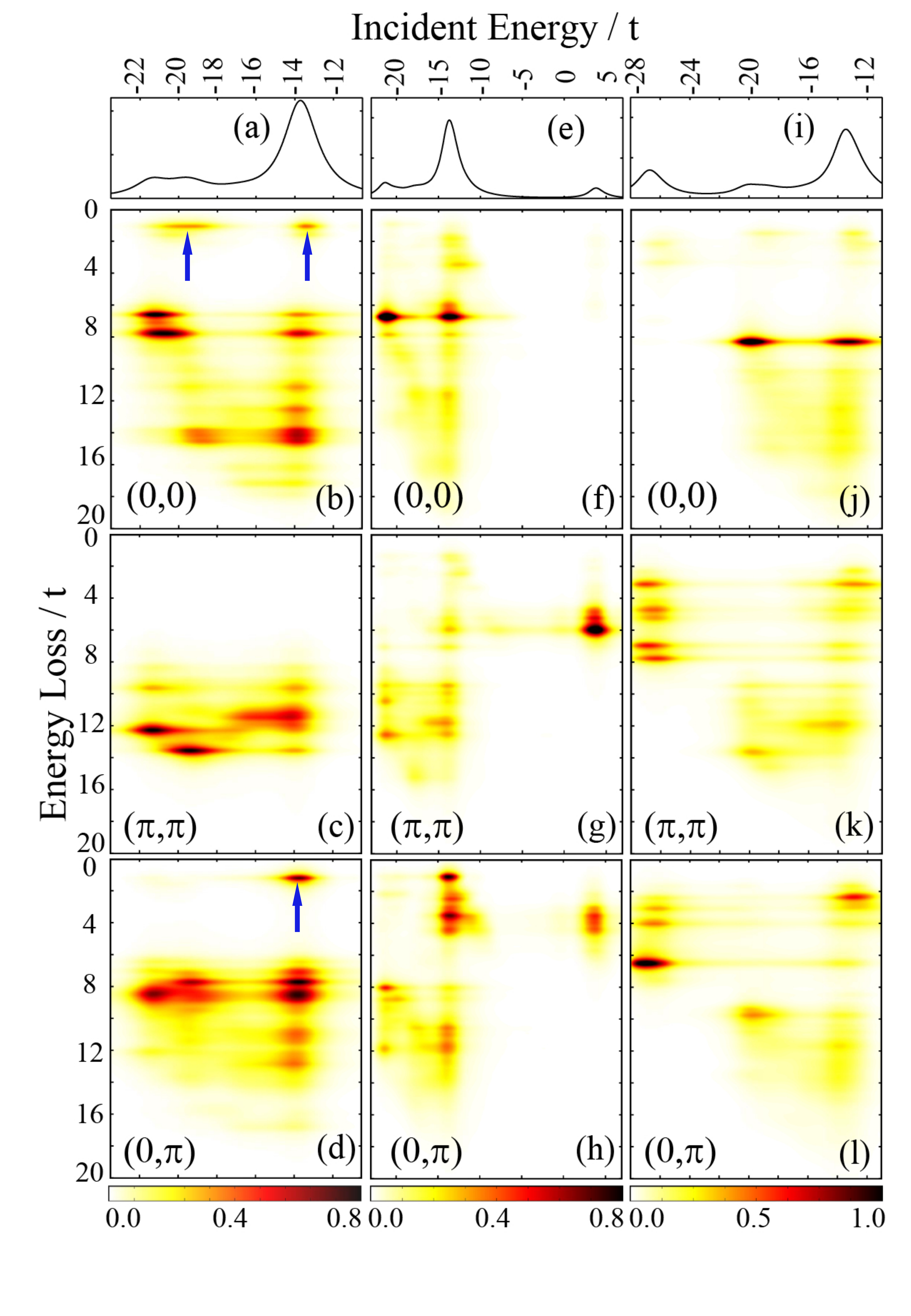}
\end{center}
\caption{(Color online) (a)-(d) XAS and momentum-dependent indirect RIXS cross-section at momentum (0,0), $(\pi,\pi)$ and $(0,\pi)$ at half-filling in the single-band Hubbard model. The filled blue arrows in (b) and (d) highlight two-magnon peaks in the RIXS cross-section. (e)-(h) the same for 12.5\% hole-doping and (i)-(l) 12.5\% electron-doping, respectively. The elastic lines of RIXS have been removed. The incident energy window is chosen so that the main XAS peaks and RIXS features display for each doping. Zero incident energy is defined as the sum of ground state energy $\mathit{E}_0$ and $\mathit{E}_{edge}$. $\mathit{E}_{edge}$ is the energy difference between the two levels in the dipole transition process, such as in Cu \textit{K}-edge RIXS case equaling $\mathit{E}_{4p}-\mathit{E}_{1s}$.}\label{fig:RIXS}
\end{figure}

\begin{figure}[t]
\begin{center}
\includegraphics[width=1.25\columnwidth]{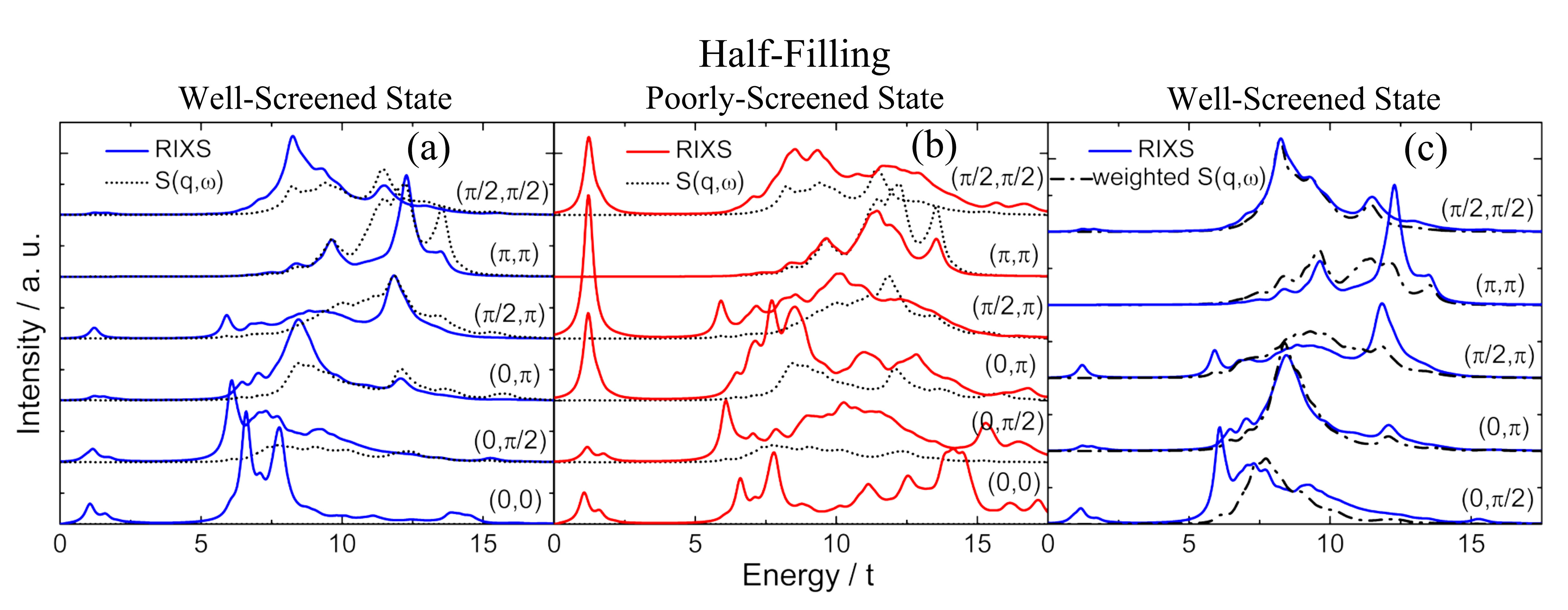}
\end{center}
\caption{(Color online) Momentum-dependent RIXS and $S(\mathbf{q},\omega)$ in the half-filled single-band Hubbard model highlighting the (a) well-screened intermediate state at incident energy $E_{in}=-21.2t$ and (b) poorly-screened intermediate state at $E_{in}=-13.85t$. (c) RIXS at the well-screened intermediate state incident energy [from (a)] vs. $S(\mathbf{q},\omega)$ multiplied by a Lorentzian prefactor given by $((\omega_{f}-\omega_{res})^2+\Gamma_{f}^2)^{-1}((\omega_{i}-\omega_{res}-|U|)^2+\Gamma_{i}^2)^{-1}$, where $\omega_{res}$ is a free parameter.\cite{YJKimRIXS} We consider equal incident/outgoing photon lifetimes ($\Gamma_i=\Gamma_f$) for fitting. Note that the elastic lines of RIXS and $S(\mathbf{q}=\mathbf{0},\omega)$ have been removed.}\label{fig:RIXS_HF}
\end{figure}

Our calculations are done on the $16B$ Betts cluster (16-site square cluster) with periodic boundary conditions. The ground state wave function is obtained using the exact diagonalization (ED) technique utilized by PARPACK(Parallel Arnoldi PACKage)\cite{PARPACK}. This algorithm is based on the implicitly restarted arnoldi method, which takes into account properly orthogonalized eigenvectors at each step of the iteration, avoiding the problem of degeneracy collapse common in most other Lanczos methods\cite{AlgBook}, such as those used in Ref.~\cite{MaekawaRIXShf,MaekawaRIXSdp} This is crucial to understanding intensities of the RIXS cross section since all intermediate states must be retained in a properly orthonormalized way. $| \Psi \rangle$ from  Eq.~3 is calculated using the bi-conjugate gradient stabilized method (BiCGSTAB)\cite{BiCGSTAB}, a variant of the bi-conjugate gradient method (BiCG). In BiCGSTAB, generalized minimal residual method (usually abbreviated GMRES) is applied after each step of BiCG in order to get rid of the irregular convergence behavior, thus obtaining faster and smoother convergence than regular BiCG. For obtaining the final spectrum, we also employed the continued fraction expansion method. 

We use the model parameters taken from Cu \textit{K}-edge RIXS in cuprate (La$_2$CuO$_4$) as an example\cite{ARPESdata}: $U=10t$, $t^{\prime}=-0.34t$, $U_c=15t$ and $\Gamma = t$. Our results will be presented in units of $t$, with $t=0.35$eV. For the indirect RIXS processes in other correlated materials, the parameter values can vary, but the key feature from correlation effect remains similar.

\section{Results}
\subsection{RIXS}

Figure \ref{fig:RIXS} shows the momentum-dependent XAS and indirect RIXS spectra for the half-filled, 12.5\% hole-doped and 12.5\% electron-doped single-band Hubbard model. Zero incident energy is defined as the ground state energy $\mathit{E}_0$ plus $\mathit{E}_{edge}$, the energy difference between the two levels in the dipole transition process. For example, $\mathit{E}_{edge}$ equals $\mathit{E}_{4p}-\mathit{E}_{1s}$ for Cu \textit{K}-edge RIXS. For the half-filling and dopings, RIXS displays strong resonances at the peaks of XAS, with the energies corresponding to the difference between the ground state and intermediate state energy in the RIXS process. 

The nature of the intermediate state can be understood in terms of how well the valence electrons screen the core-hole potential. The XAS peak at an energy $\sim-2U_c$ + $U$ ($\sim20t$ for half-filling, $\sim22t$ for 12.5\% hole-doping and $\sim20t$ for 12.5\% electron-doping) corresponds to the intermediate state with two electrons bound to the core-hole site that screen the core-hole potential. This intermediate state exhibits strong screening and is usually termed the ``well-screened" state\cite{YJKim2002_wellscreen,LuRIXS}. The XAS peak at $\sim-U_c$ ($\sim13.7t$ for half-filling, $\sim14t$ for 12.5\% hole-doping and $\sim13.5t$ for 12.5\% electron-doping) corresponds to the intermediate state where only a single electron is bound at the core-hole site, weakly screening the core-hole potential, usually termed the ``poorly-screened" state\cite{YJKim2002_wellscreen,LuRIXS}. In both processes, the electrons experience a strong ``shake-up" by the core-hole potential via screening. On the other hand, the intermediate state at an incident energy $\sim3t$ for the hole-doped or $\sim-28t$ for the electron-doped system represents the state in which the core-hole is created at an empty or a doubly-occupied site, respectively. These two states can be denoted as ``unscreened" states, where the core-hole potential does not appreciably ``shake-up" the ground state electronic system.

We next examine the energy loss dependence of RIXS spectra on resonances in Fig.~\ref{fig:RIXS}. For half-filling (Fig.~\ref{fig:RIXS}(b)-(d)), RIXS displays a high energy spectrum ranging from $\sim 5t$ up to $\sim 15t$ in energy loss for both well-screened and poorly-screened intermediate states, signaling excitations from the lower Hubbard band (LHB) to the upper Hubbard band (UHB), which is mainly controlled by Coulomb $U$. The main peaks have strong momentum dependence with the shift to higher loss energy at larger momentum transfer, due to the increase in phase space available for creation of excitations in the particle-hole continuum\cite{PHcon}. The main peaks are located at $\sim 6-8t$ for momentum $(0,0)$, and shifted to $\sim 8-9t$ at momentum $(0,\pi)$, $\sim12-14t$ at momentum $(\pi,\pi)$. This dispersion with increasing momentum at poorly- and well-screened intermediate states can be also seen in Fig.~2. In the electron- and hole-doped cases, the same charge transfer dispersion on UHB is observed also for the well- and poorly-screened intermediate states. These charge transfer features are consistent with various experiments\cite{AbbamonteRIXS,YJKimRIXS,DoringRIXS,LuRIXS} and calculations\cite{MaekawaRIXShf,MaekawaRIXSdp,ChenRIXS} carried out for the charge transfer excitations.

\subsection{Two-magnon excitations}

The RIXS spectra also exhibit distinctive structures at low energy ($\sim1-1.2t$), highlighted by blue arrows in Fig.~\ref{fig:RIXS}, emphasizing excitations of the spin degree of freedom - two-magnon excitations\cite{TwoMagnonJohnHill,TwoMagnonEllis,TwoMagnonVernay,TwoMagnonForte,TwoMagnonMagao,TwoMagnonTohyama}. This structure is found both for the poorly-screened (Fig.~2(b)) and the well-screened intermediate states (Fig.~2(a)) for the undoped system, but is significantly stronger at the poorly-screened intermediate state. In the doped system, strong intensities at low energy are also shown for the poorly-screened intermediate state (Fig.~3(b) and (e)). Only one electron is bound at the core-hole site in the poorly-screened intermediate state, thus the intermediate state wavefuncion may have a big overlap with the final state wavefunctions carrying two-magnon excitations; while for the well-screened intermediate state the two bound electrons at the core-hole site make a smaller overlap with the final states for two-magnon excitations. Previous intermediate density of states calculations also shown that the poorly-screened intermediate state has a larger weight connected to the low energy (magnetic) excitations\cite{Ahn}. 

Our data shows a strong two-magnon peak at $(\pi,0)$ momentum transfer, and no peak associated with two-magnon excitation at $(\pi,\pi)$\cite{TwoMagnonJohnHill,TwoMagnonEllis}. We also obtain two-magnon excitations at momentum transfer $(0,0)$, which are symmetry allowed by the core-hole intermediate states. Similar findings at momentum transfer $(0,0)$ have also been reported in early calculations\cite{MaekawaRIXShf,TwoMagnonVernay}. We show the two-magnon peaks at momentum transfer $(0,0)$ located at $\sim 2.7J$ (the effective exchange coupling $J = 4t^2/U$ and is $0.4t$ in our calculation), which is in agreement with the calculation taking into account the magnon-magnon interaction\cite{TwoMagnonVernay}. The two-magnon excitations exhibited in RIXS can be connected to the excitations revealed by the spin dynamical structure factor\cite{LuukRIXS}. Single-magnon excitations are forbidden in this case due to spin conservation, as we have neglected spin-orbit coupling. 

Our calculations are consistent with the Cu \textit{K}-edge experiments of the magnetic excitations taken on La$_2$CuO$_4$ for the observation of strong peak at momentum transfer $(\pi,0)$ and no weight at momentum transfer $(\pi,\pi)$\cite{TwoMagnonJohnHill,TwoMagnonEllis}. No experimental confirmation of the two-magnon feature at momentum $(0,0)$ has been reported, although recent improvements in resolution may resolve this issue. The two-magnon excitation is significant at the poorly-screened intermediate state compared to the others providing a focus to the fact that particular elementary excitation may be highlighted when tuning to a proper incident photon energy dictated by the character of the intermediate state.

\begin{figure}[t]
\begin{center}
\includegraphics[width=1.25\columnwidth]{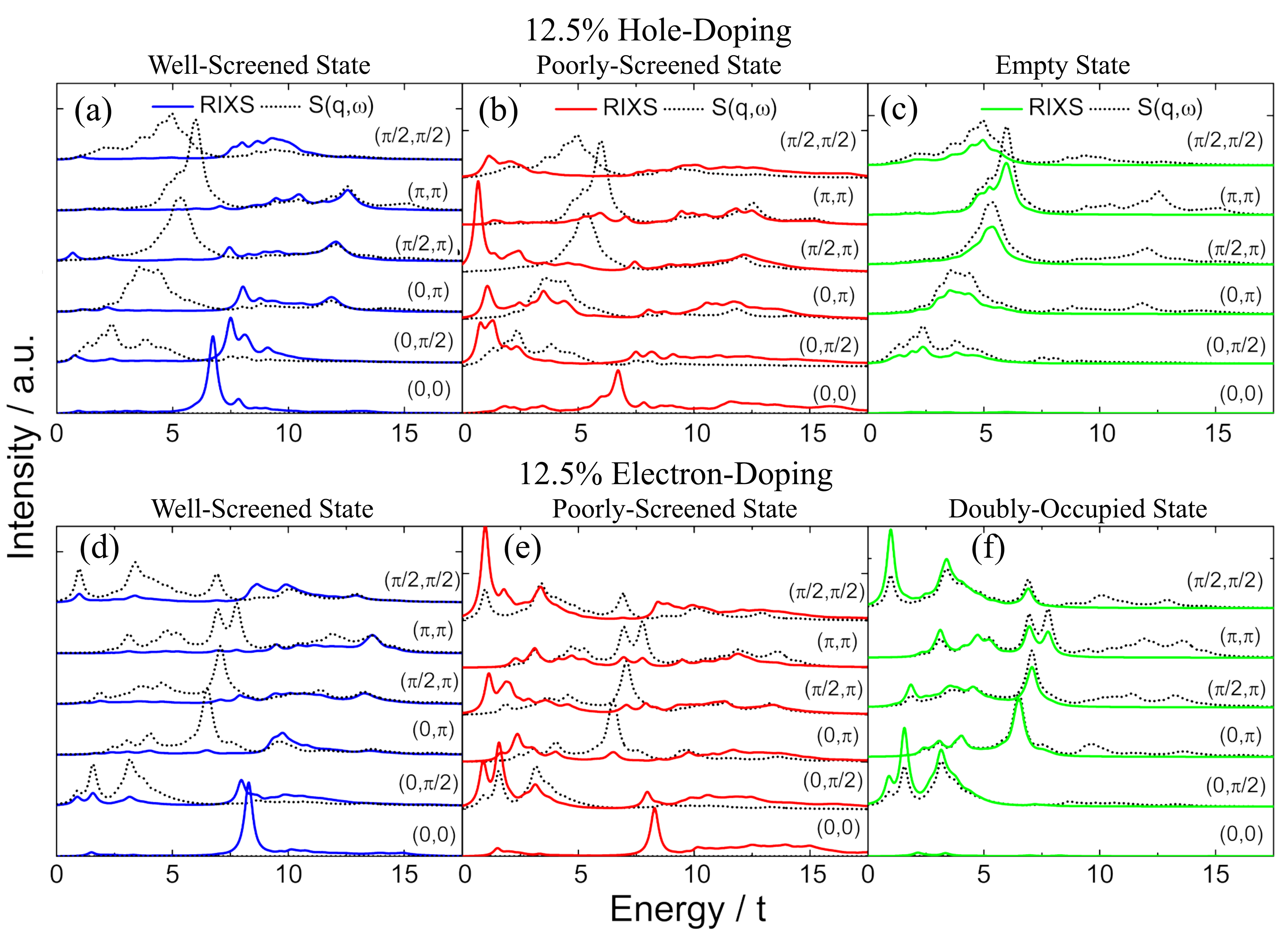}
\end{center}
\caption{(Color online) Momentum dependent RIXS and $S(\mathbf{q},\omega)$ at 12.5\% hole-doping [(a)$E_{in}=-21.2t$ (b)$E_{in}=-14.2t$ (c)$E_{in}=3.78t$] and 12.5\% electron-doping [(d)$E_{in}=-19.75t$ (e)$E_{in}=-13.33t$ (f)$E_{in}=-26.88t$] for single-band Hubbard model. We note that the elastic lines of RIXS and $S(\mathbf{q}=\mathbf{0},\omega)$ have been removed. Blue, red and green lines represent RIXS spectra with the intermediate state having the core-hole well-screened, poorly-screened and created at an empty/doubly-occupied site, respectively. }\label{fig:RIXS_Doped}
\end{figure}

\subsection{RIXS vs. $S(\mathbf{q},\omega)$}
To make comparisons between RIXS and $S(\mathbf{q},\omega)$, in Fig.~\ref{fig:RIXS_HF} we plot the RIXS spectra for the well- and poorly-screened intermediate states, compared to $S(\mathbf{q},\omega)$, at half-filling. When $\omega$ $>5t$, $S(\mathbf{q},\omega)$ (black dotted) and the RIXS spectra (colored solid) generally span the same energy range at a given momentum transfer, and both shift to higher energy at larger momentum transfer. Nonetheless, their overall intensities are very different: at momentum ($0, \pi/2$), ($0,\pi$) and ($\pi,\pi/2$) the RIXS spectra have more weight at lower energy and a lower threshold; at momentum ($\pi, \pi$) and ($\pi/2, \pi/2$) the RIXS spectra and $S(\mathbf{q},\omega)$ exhibit different main peaks and weights. Another striking difference lies in the $(0,0)$ momentum transfer. For $S(\mathbf{q},\omega)$ no energy loss excitations are allowed, since the charge operator $\rho_{\mathbf{q}}^d$ at $\mathbf{q}=\mathbf{0}$ (see Eq.~3) becomes the total number of electron operator that commutes with the Hamiltonian, so it cannot give any inelastic response; however, various excitations exist in RIXS at zero momentum and there is no such symmetry that forbids the excitations. 

To investigate the conjecture that RIXS and $S(\mathbf{q},\omega)$ can be connected by a simple resonant prefactor, we compare RIXS at the well-screened intermediate state with $S(\mathbf{q},\omega)$ multiplied by a fitted Lorentzian prefactor as shown in Fig.~2(c). These results show that these two spectra only agree at momentum ($0, \pi$) and ($\pi/2, \pi/2$); therefore, a simple Lorentzian prefactor cannot in general address the complicated differences between the spectra. The above analysis from half-filling also applies to other dopings for the poorly- and well-screened intermediate states, as shown in Fig.~\ref{fig:RIXS_Doped}(a)-(b) and(d)-(e).

In some limited cases RIXS and $S(\mathbf{q},\omega)$ agree well with one another, specifically for charge excitations within the LHB. This happens for the ``unscreened" intermediate states [Fig.~\ref{fig:RIXS_Doped}(c) and (f)], in which the core-hole potential does not significantly ``shake-up" the electrons, and the effect of the core-hole potential can be neglected or is less prominent. In this case, the complicated four-particle RIXS process could be simplified to the two-particle $S(\mathbf{q},\omega)$ process. Our results show that if the core-hole ``shakes-up" the electrons when $U_c$ is comparable to the other energy scales, RIXS deviates from $S(\mathbf{q},\omega)$. If there is a lack of this ``shake-up" or core-hole potential screening, RIXS may be approximated by $S(\mathbf{q},\omega)$ with a resonant prefactor. Previous theoretical treatment of connecting indirect RIXS the same as $S(\mathbf{q},\omega)$ works at weak $U_c$ for the series expansion\cite{AbbamonteRIXS} and strong and weak $U_c$ for the ultrashort core-hole life expansion\cite{VDBRIXS,LuukRIXS}. These schemes do not describe the system with an intermediate $U_c$ and fail to address the RIXS cross-section difference associated with the different intermediate state configuration. We note that typically $U_c$ is comparable to other energy scales in the problem mostly connected with the Coulomb energy scale for $d$-electrons, and therefore an expansion may have no small parameter. However, if the intermediate state does not substantially involve these energy scales, then the approach used in Refs 6,7 would be appropriate. This is in agreement with our results: when the intermediate state involves a core-hole created at an empty state, the core-hole interaction is ineffective, meaning that the RIXS cross section can be then related to the charge response. In general however, the role of the core hole interaction must be retained for a faithful representation of the RIXS response.

Experimental evidences of the core-hole potential screening on the indirect RIXS and the comparison to $S(\mathbf{q},\omega)$ have been found on various cuprate materials\cite{YJKimRIXS}. For those cuprates with less screening channels, such as 1D cuprates, RIXS shows similar response as  $S(\mathbf{q},\omega)$; while for 2D cuprates with more screening channels the deviations of the two spectra look more obvious. 

\section{Summary and discussions}

In summary, we have employed exact diagonalization to study the resonance profile for indirect RIXS over a wide range of dopings and incident energies in a simple correlated model. The RIXS spectra display complex incident energy dependence in the resonant profile that offers rich information besides a simple boost of intensity. Controversial to previous understanding that RIXS is $S(\mathbf{q},\omega)$ in the charge transfer channel, we found that RIXS could only be considered as  $S(\mathbf{q},\omega)$ when the screening of core-hole potential at the corresponding intermediate state is not important.  Moreover, our results show strong two-magnon excitations for poorly-screened intermediate state. We conclude that at different incident energies, certain subdominant characters of final excitations can therefore be identified with a corresponding resonant enhancement in the RIXS profile\cite{ChenRIXS}. One could thus explore particular elementary excitations by focusing on certain incident energies in the resonant profile.

We note that to make a better comparison to experimental Cu \textit{K}-edge RIXS or XAS data, one needs to do a convolution with the Cu 4p density of states for the incident photon energy\cite{ChenRIXS}. For experimental indirect RIXS data, the spectra on one incident energy may not be related to a single intermediate state configuration, since the convolution may mix the contributions from different intermediate states. Thus for studying a certain elementary excitation in RIXS, such as two-magnon or charge dynamical structure factor, one can focus on the incident phonon energy which has the biggest weight connecting the specific intermediate state through the convolution. One may also do a deconvolution for the analysis of the experimental data. We emphasis that our purpose for this single-band Hubbard model study was to investigate how the nature of a particular intermediate electronic structure affects the RIXS spectra, and look into the momentum dependent RIXS spectra at different intermediate states. More complicated multi-band models may also be needed to address more elementary excitations such as the \textit{dd} excitations, although less momentum points are accessible since we can only study smaller cluster as more bands are included in the Hamiltonian. For cuprate materials in reality, the intermediate state also has ligand character which beyond our down-folded model, but the intermediate state contribution to RIXS spectra in terms of the core-hole potential screening remains an important observation\cite{YJKimRIXS}. Our model also well addresses the two-magnon excitaions and charge transfer dispersion as in experiments for cuprates. 

As the single-band Hubbard model carries the key character of correlated materials, our conclusion is not limited in cuprates but could also be generalized to all correlated materials in understanding indirect RIXS response on general forms. When the core-hole potential is comparable to other energy scales, the importance of the intermediate state configuration cannot be ignored and the incident energy dependence must be considered. The screening of the core-hole potential is a key factor for selecting the best incident photon energy to study certain elementary excitations.

\ack
We would like to thank J. P. Hill for valuable discussions. This work was supported at SLAC and Stanford University by the U.S. Department of Energy, Office of Basic Energy Sciences, Division of Materials Science and Engineering, under Contract No.~DE-AC02-76SF00515 and by the Computational Materials and Chemical Sciences Network (CMCSN) under Contract No.~DE-FG02-08ER46540. C.~J.~Jia is also supported by the Stanford Graduate Fellows in Science and Engineering. A portion of the computational work was performed using the resources of the National Energy Research Scientific Computing Center (NERSC) supported by the U.S. Department of Energy, Office of Science, under Contract No.~DE-AC02-05CH11231.

\section*{References}

\bibliographystyle{iopart-num}
\bibliography{rixs}

\providecommand{\newblock}{}
\begin{thebibliography}{10}
\expandafter\ifx\csname url\endcsname\relax
  \def\url#1{{\tt #1}}\fi
\expandafter\ifx\csname urlprefix\endcsname\relax\def\urlprefix{URL }\fi
\providecommand{\eprint}[2][]{\url{#2}}

\bibitem{KotaniReview}
Kotani A and Shin S 2001 {\em Rev. Mod. Phys.\/} {\bf 73} 203

\bibitem{TomRIXSReview}
Ament L~J~P, van Veenendaal M, Devereaux T~P, Hill J~P and van~den Brink J 2011
  {\em Rev. Mod. Phys\/} {\bf 83} 705

\bibitem{NatPhyRIXS}
Le~Tacon M, G~Ghiringhelli Chaloupka J, Moretti~Sala M~Hinkov V, Haverkort M~W,
  Minola M, Bakr M, Zhou K~J, Blanco-Canosa S, Monney C, Song Y~T, Sun G~L, Lin
  C~T, De~Luca G~M, Salluzzo M, Khaliullin G, Schmitt T, Braicovich L and
  Keimer B 2011 {\em Nat. Phys.\/} {\bf 7} 725

\bibitem{Tom2007Review}
Devereaux T~P and Hackl R 2007 {\em Rev. Mod. Phys.\/} {\bf 79} 175

\bibitem{AbbamonteRIXS}
Abbamonte P, Burns C~A, Isaacs E~D, Platzman P~M, Miller L~L, Cheong S~W and
  Klein M~V 1999 {\em Phys. Rev. Lett.\/} {\bf 83} 860

\bibitem{VDBRIXS}
van~den Brink J and Veenendaal M~v 2006 {\em Europhys. Lett.\/} {\bf 73} 121

\bibitem{LuukRIXS}
Ament L~J~P, Forte F and van~den Brink J 2007 {\em Phys. Rev. B\/} {\bf 75}
  115118

\bibitem{YJKim2004}
Kim Y~J, Hill J~P, Benthien H, Essler F~H~L, Jeckelmann E, Choi H~S, Noh T~W
  and Motoyama N 2004 {\em Phys. Rev. Lett.\/} {\bf 92} 137402

\bibitem{YJKim2007}
Kim Y~J, Hill J~P, Wakimoto S, Birgeneau R~J, Chou F~C, Motoyama N, Kojima K~M,
  Uchida S, Casa D and Gog T 2007 {\em Phys. Rev. B\/} {\bf 76} 155116

\bibitem{Grenier}
Grenier S, Hill J~P, Kiryukhin V, Ku W, Kim Y~J, Thomas K~J, Cheong S~W, Tokura
  Y, Tomioka Y, Casa D and Gog T 2005 {\em Phys. Rev. Lett.\/} {\bf 94} 047203

\bibitem{YJKimRIXS}
Kim J, Ellis D~S, Zhang H, Kim Y~J, Hill J~P, Chou F~C, Gog T and Casa D 2009
  {\em Phys. Rev. B\/} {\bf 79} 094525

\bibitem{TwoMagnonEllis}
Ellis D~S, Kim J, Hill J~P, Wakimoto S, Birgeneau R~J, Shvyd'ko Y, Casa D, God
  T, Ishii K, Ikeuchi K, Paramekanti A and Kim Y~J 2010 {\em Phys. Rev. B\/}
  {\bf 81} 085124

\bibitem{TwoMagnonForte}
Forte F, J~P~Ament L and van~den Brink J 2008 {\em Phys. Rev. B\/} {\bf 77}
  134428

\bibitem{TwoMagnonJohnHill}
Hill J~P, Blumberg G, Kim Y~J, Ellis D~S, Wakimoto S, Birgeneau R~J, Komiya S,
  Ando Y, Liang B, Greene R~L, Casa D and God T 2008 {\em Phys. Rev. Lett.\/}
  {\bf 100} 097001

\bibitem{KramersHeisenberg}
Kramers H~A and Heisenberg W 1925 {\em Z. Phys.\/} {\bf 31} 681

\bibitem{PARPACK}
Lehoucq R~B, Sorensen D~C and Yang C 1998 {\em ARPACK Users Guide: Solution of
  Large-Scale Eigenvalue Problems with Implicitly Restarted Arnoldi Methods.\/}
  (Philadelphia: SIAM)

\bibitem{AlgBook}
Weisse A and Fehske H 2008 {\em Computational Many Particle Physics\/}
  (Springer)

\bibitem{MaekawaRIXShf}
Tsutsui K, Tohyama T and Maekawa S 1999 {\em Phys. Rev. Lett.\/} {\bf 83} 3705

\bibitem{MaekawaRIXSdp}
Tsutsui K, Tohyama T and Maekawa S 2003 {\em Phys. Rev. Lett.\/} {\bf 91}
  117001

\bibitem{BiCGSTAB}
Van~der Vorst H~A 1992 {\em SIAM J. on Scientific and Statistical Computing\/}
  {\bf 13} 631

\bibitem{ARPESdata}
Kim C, While P~J, Shen Z~X, Tohyama T, Shibata Y, Maekawa S, Wells B~O, Kim
  Y~J, Birgeneau R~J and Kastner M~A 1998 {\em Phys. Rev. Lett.\/} {\bf 80}
  4245

\bibitem{YJKim2002_wellscreen}
Kim Y~J, Hill J~P, Burns C~A, Wakimoto S~R, Birgeneau J, Casa D, Gog T and
  Venkataraman C~T 2002 {\em Phys. Rev. Lett.\/} {\bf 89} 177003

\bibitem{LuRIXS}
Lu L, Hancock J~N, Chabot-Couture G, Ishii K, Vajk O~P, Yu G, Mizuki J, Casa D,
  Gog T and Greven M 2006 {\em Phys. Rev. B\/} {\bf 74} 224509

\bibitem{PHcon}
Devereaux T~P, McCormack G~E~D and Freericks J~K 2003 {\em Phys. Rev. Lett.\/}
  {\bf 90} 067402

\bibitem{DoringRIXS}
Doring G, Sternemann C, Kaprolat A, Mattila A, Hamalainen K and Schulke W 2004
  {\em Phys. Rev. B\/} {\bf 70} 085115

\bibitem{ChenRIXS}
Chen C~C, Moritz B, Vernay F, Hancock J~N, Johnston S, Jia C, Ghabot-Couture G,
  Greven M, Elfimov G~A, Sawatzky G~A and Devereaux T~P 2010 {\em Phys. Rev.
  Lett.\/} {\bf 105} 177401

\bibitem{TwoMagnonVernay}
Vernay F~H, Gingras J~P and Devereaux T~P 2007 {\em Phys. Rev. B\/} {\bf 75}
  020403(R)

\bibitem{TwoMagnonMagao}
Nagao T and Igarashi J~i 2007 {\em Phys. Rev. B\/} {\bf 75} 214414

\bibitem{TwoMagnonTohyama}
Tohyama T, Onodera H, Tsutsui K and S M 2002 {\em Phys. Rev. Letts.\/} {\bf 89}
  257405

\bibitem{Ahn}
Ahn K~H, Fedro A~J and van Veenendaal M 2009 {\em Phys. Rev. B\/} {\bf 79}
  045103

\end{thebibliography}

\end{document}